\documentstyle[aps,preprint,epsf]{revtex}

\begin{document}

\draft
\tightenlines

\title{Mechanisms of positron annihilation on molecules}

\author{G. F. Gribakin\cite{affil}}

\address{School of Physics, University of New South Wales,
Sydney 2052, Australia}

\maketitle

\begin{abstract}

The aim of this work is to identify the mechanisms responsible for very large
rates and other peculiarities observed in low-energy positron annihilation
on molecules. The two mechanisms considered are:
(i) Direct annihilation of the incoming positron with one of the molecular
electrons. This mechanism dominates for atoms and small molecules. I show that
its contribution to the annihilation rate can be related to the positron
elastic scattering cross section. This mechanism is characterized by strong
energy dependence of $Z_{\rm eff}$ at small positron
energies and high $Z_{\rm eff}$ values (up to $10^3$) for room temperature
positrons, if a low-lying virtual level or a weakly bound state exists for the
positron. (ii) Resonant annihilation, which takes place when the positron
undergoes resonant capture into a vibrationally excited quasibound state of the
positron-molecule complex. This mechanism dominates for larger molecules
capable of forming bound states with the positron. For this mechanism
$Z_{\rm eff}$ averaged over some energy interval, e.g., due to thermal positron
energy distribution, is proportional to the level density of the
positron-molecule complex, which is basically determined by the spectrum of
molecular vibrational states populated in the positron capture. The resonant
mechanism can produce very large annihilation rates corresponding to
$Z_{\rm eff}\sim 10^8$. It is highly sensitive to molecular structure and
shows a characteristic $\varepsilon ^{-1/2}$ behaviour of $Z_{\rm eff}$ at
small positron energies $\varepsilon $. The theory is used to analyse
calculated and measured $Z_{\rm eff}$ for a number of atoms and molecules.

\end{abstract}

\newpage
%*****************************************************************************

\section{Introduction}\label{sec:intro}

The aim of this work is to develop the framework for the description of
low-energy positron annihilation on molecules, and to analyse its two main
mechanisms: direct and resonant annihilation. There are a number of remarkable
phenomena associated with this process: very large annihilation rates
\cite{Paul:63,Heyland:82,Surko:88}, high sensitivity of the rates to small
changes in the molecular structure \cite{Iwata:95}, large
ionization-fragmentation cross sections for organic molecules at
sub-Ps-threshold positron energies \cite{Hulett:93}, and rapid increase
of the fragmentation and annihilation rates towards small positron energies
\cite{JunXu:94,Iwata:99}. In spite of decades of study, there is no consistent
physical picture or even general understanding of these processes, and there
have been very few calculations \cite{daSilva:96}, which leaves too much room
for speculations \cite{Laricchia:97}. My main objective is to
consider real mechanisms of positron annihilation on molecules, describe their
characteristic features, make estimates of the corresponding annihilation
rates, and formulate the terms in which positron-molecule annihilation should
be described and analysed. In recent work \cite{Iwata:99}, Iwata {\em et al.}
describe new experiments to study positron annihilation on molecules. Some of
these experiments test specific features of the annihilation processes
described in the present paper. Though some aspects of the experimental work
are discussed here, futher details and comparison with theory and various
models of positron annihilation can be found in Ref. \cite{Iwata:99}.

The annihilation rate $\lambda $ for positrons in a molecular or atomic gas
is usually expressed in terms of a dimensionless parameter $Z_{\rm eff}$:
\begin{equation}\label{eq:zeff}
\lambda =\pi r_0^2 c Z_{\rm eff} n,
\end{equation}
where $r_0$ is the classical radius of the electron, $\pi r_0^2 c$ is the
non-relativistic spin-averaged rate of electron-positron annihilation into two
$\gamma $ quanta, and $n$ is the number density of molecules \cite{note}.
Equation (\ref{eq:zeff}) implies that $Z_{\rm eff}$ is the effective number of
target electrons contributing to the annihilation process. In terms of the
annihilation cross section $\sigma _a$ the rate is $\lambda =\sigma _anv$, so
by comparison with Eq. (\ref{eq:zeff}), we have
\begin{equation}\label{eq:siganzeff}
\sigma _a= \pi r_0^2 c Z_{\rm eff}/v,
\end{equation}
where $v$ is the positron velocity. Accordingly, the spin-averaged cross
section of annihilation of a non-relativistic positron on a single electron
corresponds to $Z_{\rm eff}=1$, see e.g. \cite{Akhiezer:65}. If the
annihilation occurs during binary positron-molecule collisions, as in the
experiments of the San Diego group \cite{Surko:88,Iwata:95,Iwata:PhD} who use a
positron trap and work at low gas densities, the parameter $Z_{\rm eff}$ is
independent of the density. It characterizes the annihilation of a positron
on a single molecule.

One could expect that $Z_{\rm eff}$ is comparable to the number of electrons
$Z$ in an atom or molecule. Moreover, low-energy positrons do
not penetrate deep into the atom, and annihilate most probably with the
valence electrons only. However, even for hydrogen $Z_{\rm eff}=8$ at low
energies \cite{Humberston:72}. This is a manifestation of correlation effects.
The most important of them is polarization of the atom by the positron and,
as a result, an attractive $-\alpha e^2/2r^4$ positron-atom potential,
$\alpha $ being the atomic dipole polarizability. An additional short-range
contribution to the positron-atom attraction comes from virtual Ps formation,
i.e., hopping, or rather, tunneling of an electron between the atomic ion and
the positron. The electron density on the positron is also enhanced due to the
Coulomb attraction between them. These effects make atomic $Z_{\rm eff}$ large,
e.g., $Z_{\rm eff}=401$ for room temperature positrons on Xe \cite{Murphy:90}.

\begin{figure}[t]
\vspace{-100pt}
\epsfxsize=15.000cm
\centering\leavevmode\epsfbox{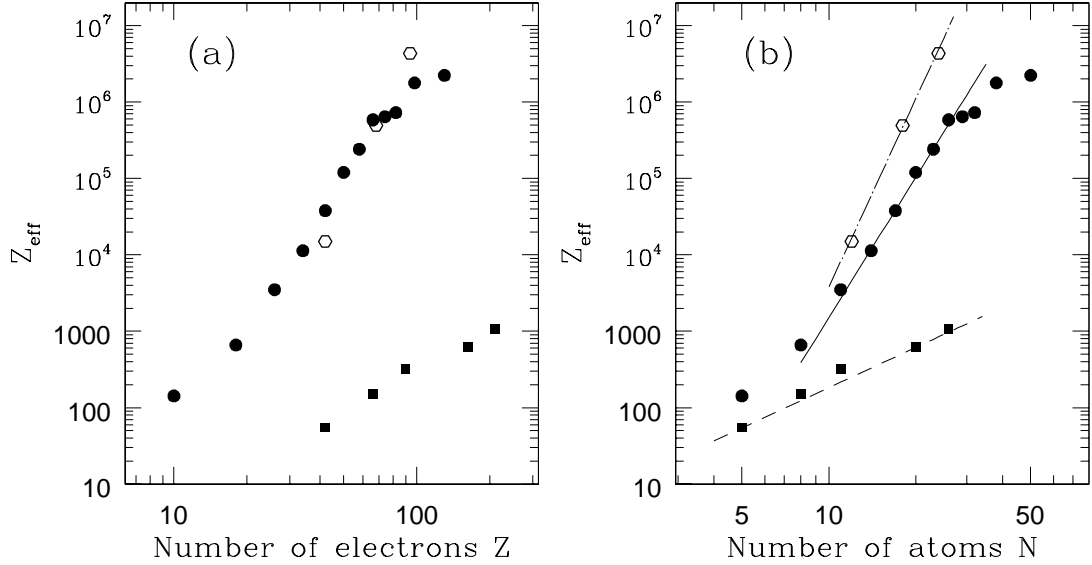}
\vspace{-90pt}
\caption{Annihilation rates $Z_{\rm eff}$ for alkanes, C$_n$H$_{2n+2}$
(solid circles, $n=1$--10, 12, and 16), perfluorinated alkanes,
C$_n$F$_{2n+2}$ (solid squares, $n=1$--3, 6, and 8) and aromatic hydrocarbons,
benzene, naphthalene, and antracene, C$_n$H$_{n/2+3}$ (open hexagons,
$n=6$, 10, and 14), as functions of the number of electrons in the molecule
$Z$ (a), and number of atoms $N$ (b). Data are taken from Ref.
\protect \cite{Iwata:PhD}, Tables B1, 4.3, 4.9 and 4.11 (see also
\protect \cite{Iwata:95}). Also shown are power-law fits for
alkanes,
% $Z_{\rm eff}\propto Z^{5.5}$, and
$Z_{\rm eff}\propto N^{6.1}$ (solid line), perfluorinated alkanes,
% $Z_{\rm eff}\propto Z^{1.8}$, and
$Z_{\rm eff}\propto N^{1.75}$ (dashed line), and aromatic hydrocarbons,
% $Z_{\rm eff}\propto Z^{7.0}$, and
$Z_{\rm eff}\propto N^{8.2}$ (dot-dashed line).}
\label{fig:Zeffmol}
\end{figure}

Even compared with this large number, annihilation rates for low-energy
(room temperature) positrons on polyatomic molecules are huge. They
increase very rapidly with the molecular size, and depend strongly on the
chemical composition of the molecules, see Fig.~\ref{fig:Zeffmol}. This has
been known for quite a while, after early measurements for CCl$_4$,
$Z_{\rm eff}=2.2\times 10^4$ \cite{Paul:63}, butane, $1.5\times 10^4$
\cite{Heyland:82}, and $Z_{\rm eff}$ ranging between $10^4$ and $2\times 10^6$
for large alkanes C$_n$H$_{2n+2}$, $n=$4--16 \cite{Leventhal:90} (see also
\cite{Iwata:95}). The largest $Z_{\rm eff}$ values measured so far are
$4.3\times 10^6$ for antracene C$_{14}$H$_{10}$ \cite{Murphy:91} and
$7.5\times 10^6$ for sebacic acid dimethyl ester C$_{12}$H$_{22}$O$_4$
\cite{Leventhal:90}. Thus, while $Z_{\rm eff}$ up to five orders of magnitude
greater than $Z$ have been observed, the physical processes responsible for
these anomalously large annihilation rates have not been really understood.
In other words, if the observed $Z_{\rm eff}$ are parametrically large,
compared to the number of available electrons, then what are the parameters
that determine large annihilation rates for positrons on molecules?

In this work I consider two basic mechanisms of positron-molecule annihilation.
The first mechanism is {\em direct annihilation} of the incoming positron with
one of the molecular electrons. The contribution of
this mechanism to the annihilation rate is proportional to the number of
valence electrons available for annihilation. It can be enhanced by the
positron-molecule interaction which distorts the positron wave. In
particular,  the positron density in the vicinity of the molecule increases
greatly if a low-lying virtual state ($\varepsilon _0>0$) or a weakly bound
level ($\varepsilon _0<0$) exists for the $s$-wave positron. In this case
$Z_{\rm eff}^{\rm (dir)}\propto 1/(\varepsilon + |\varepsilon _0|)$
for small positron energies $\varepsilon \lesssim |\varepsilon _0|$
\cite{Goldanskii:64,Dzuba:93,Dzuba:96}. This type of enhancement is responsible
for large $Z_{\rm eff}$ values observed in heavier noble gas atoms, where
successively lower virtual levels exist for the positron ($Z_{\rm eff}=33.8$,
90.1 and 401 for Ar, Kr and Xe, respectively \cite{Iwata:95,Murphy:90}).
This understanding is confirmed by the temperature dependences of the
annihilation rates measured for the noble gases in \cite{Kurz:96}. Note that
for room-temperature positrons, $\varepsilon \sim k_{\rm B}T$, even for
$\varepsilon _0 \rightarrow 0$ the size of the enhancement due to
virtual/weakly bound states is limited.

The second mechanism is {\em resonant annihilation}. By this I mean a
two-stage process. The positron is first captured into a Feshbach-type
resonance, where positron attachment is accompanied by excitation of some
molecular degrees of freedom. Such process is well known for
electrons \cite{Christo:84}. The positron in the quasi bound state then
annihilates with a molecular electron. Enhancement of annihilation due to a
single resonance was considered theoretically
in \cite{Smith:70,Ivanov:86}. The possibility of forming such
resonances by excitation of the vibrational degrees of freedom of molecules
was proposed by Surko {\em et al.} \cite{Surko:88} to explain high
annihilation rates and their strong dependence on the molecular size observed
for alkanes. It was also considered in relation to the problem of fragmentation
of molecules by positron annihilation \cite{Crawford:94}. However, its
contribution to the annihilation have never been properly evaluated.
To make this mechanism work for low-energy positrons one must assumed that
positrons can form bound states with large neutral molecules, i.e., the
positron affinity of the molecule is positive, $\varepsilon _A>0$
\cite{Surko:88}. The capture is then possible if the energy of the
incoming positron is in resonance with the vibrationally excited state of the
positron-molecule complex \cite{note1}. The density of
the vibrational excitation spectrum of this complex can be high, even if the
excitation energy supplied by positron binding,
$E_v=\varepsilon _A+\varepsilon $, is only few tenths of an eV
(it is reasonable to assume that the presence of the positron does not change
the vibrational spectrum of the molecule by too much). For positrons with
thermal Maxwellian energy distribution the contribution of the resonant
annihilation mechanism averaged over a number of resonances
$Z_{\rm eff}^{\rm (res)}$ is observed. The magnitude of
$Z_{\rm eff}^{\rm (res)}$ is determined by three parameters of the
positron-molecule
resonant states: their annihilation width $\Gamma _a$, the autodetachment
width $\Gamma _c$, which also determines the probability of positron
capture, and the level density $\rho (E_v)$ of the positron-molecule resonant
states populated in positron capture. The magnitude of $\Gamma _a$
for positron-molecule bound states is comparable to the spin-averaged
annihilation width of the Ps atom ($\Gamma _a/\hbar \sim 5\times 10^{-10}$~s).
Note that $\Gamma _a$ does not increase with the size of the molecule, because
the increase in the number of electrons is accompanied by thinning of the
positron density in the (quasi)bound positron-molecule state. It turns
out (see Sec. \ref{sec:anmech}) that for $\Gamma _c \gg \Gamma _a$ the
magnitude of $Z_{\rm eff}^{\rm (res)}$ is simply proportional to $\rho (E_v)$.
This density increases rapidly with the size of the molecule,
$\rho (E_v)\propto (N_v)^n$, where $N_v$ is the number of vibrational modes,
$n\sim \varepsilon _A/\omega $ is the effective number of vibrational quanta
excited in positron capture, and $\omega $ is a typical molecular vibrational
frequency. Thus, the resonant annihilation mechanism can explain the rapid
increase of $Z_{\rm eff}$ with the size of the molecule shown in
Fig.~\ref{fig:Zeffmol}. Moreover, my estimates show that for thermal positrons
$Z_{\rm eff}^{\rm (res)}$ up to $10^8$ could be observed.

A necessary condition for the resonant annihilation to occur is the
existence of positron-molecule bound states. Until recently there was almost
no positive information about the possibility of positron binding to
neutral atomic species. The experimental results and their interpretation
by Surko {\em et al.} \cite{Surko:88} could be viewed as the strongest,
albeit indirect, evidence of positron binding to large molecules. This
situation has changed now. Many-body theory calculations of Dzuba {\em et al.}
\cite{Dzuba:95} indicated strongly that positrons can be bound by Mg, Zn, Cd,
and Hg and, possibly, many other atoms. Recently the variational calculations
of Ryzhikh and Mitroy proved rigorously that positrons form bound states with
Li atoms, and demonstrated that bound states also exist for Na, Be, Mg, Zn,
Cu and Ag \cite{Ryzhikh:97}. Molecules are much larger potential wells
for the positron, and it seems natural that many of them should be capable of
binding positrons.

Ideas about different mechanisms in positron-molecule annihilation have been
discussed earlier in a number of theoretical \cite{Goldanskii:64,Ivanov:86}
and experimental \cite{Surko:88,Smith:70} works. However, there is a need
to re-examine this question using a unified approach to the annihilation
mechanisms, and define clearly the physical variables which determine the
observed annihilation rates. The latter is especially important for the
present work which aims to provide understanding of a whole variety of
phenomena, including the origins of the high values of $Z_{\rm eff}$ for
molecules and their dependence on the chemical composition and positron
energy.

%****************************************************************************
\section{Annihilation mechanisms}
\label{sec:anmech}

In this section a derivation of the positron annihilation rate within a
standard scattering theory formalism is presented. I show how to estimate the
contributions of the direct and resonant mechanisms, and examine specific
features of these mechanisms.

\subsection{General expressions}\label{subsec:gen}

The effective number of electrons $Z_{\rm eff}$ related to the annihilation
rate through Eq. (\ref{eq:zeff}) is determined by the positron density
on the electrons
\begin{equation}\label{eq:zeffpsi}
Z_{\rm eff}=\int \sum _{i=1}^Z\delta ({\bf r}-{\bf r}_i)|\Psi _{\bf k}
({\bf r}_1,\dots ,{\bf r}_Z,{\bf r})|^2d{\bf r}_1\dots d{\bf r}_Zd{\bf r}~,
\end{equation}
where $Z$ is the number of target electrons, ${\bf r}_i$ and ${\bf r}$ are
the coordinates of the electrons and positron, respectively, and
$\Psi _{\bf k}({\bf r}_1,\dots ,{\bf r}_Z,{\bf r})$ is the total wave function
of the system. It describes scattering of the positron with initial momentum
${\bf k}$ from the atomic or molecular target in the ground state $\Phi _0$,
and is normalized as
\begin{equation}\label{eq:asymp}
\Psi _{\bf k}({\bf r}_1,\dots ,{\bf r}_Z,{\bf r})\simeq \Phi _0
({\bf r}_1,\dots ,{\bf r}_Z)e^{i{\bf kr}}\quad (r\gg R_a),
\end{equation}
where $R_a$ is the radius of the target (atomic units are used throughout).
Note that for molecular targets $\Psi _{\bf k}$ and $\Phi _0$ should,
strictly speaking, depend on the nuclear coordinates as well.

Let us first assume that the electron-positron degrees of freedom are
completely decoupled from the nuclear motion. The scattering wave function
is then determined by the positron interaction with the charge
distribution of the ground-state target and electron-positron correlation
interaction (polarization of the target, virtual Ps formation, etc.). Let us
denote the corresponding wave function $\Psi ^{(0)}_{\bf k}$. At positron
energies of a few electron Volts the molecule can be excited electronically,
and the positron may find itself trapped in electronically excited Feshbach
resonance states. This may result in rapid resonant energy dependence of the
$\Psi ^{(0)}_{\bf k}$ wave function. However, at small sub-eV or
room-temperature positron energies electron excitations cannot be produced, and
$\Psi ^{(0)}_{\bf k}$ behaves smoothly. On the other hand, if the positron
affinity  of a molecule is positive, the system `molecule\,$+$\,positron' is
capable of forming a stable ``positronic ion'', whose lifetime is only limited
by positron annihilation. This system will also have a number of excited bound
states $\Phi _\nu $ corresponding to vibrational excitations of the
positron-molecule complex. Their typical energies are of the order of 0.1 eV
and smaller, as determined by the vibrational spectrum of the molecule.

If we now turn on the coupling $V$ between the electron-positron and nuclear
degrees of freedom the total scattering wave function will be given by
\begin{equation}\label{eq:totpsi}
|\Psi _{\bf k}\rangle =|\Psi ^{(0)}_{\bf k}\rangle +\sum _\nu
\frac{|\Phi _\nu \rangle \langle \Phi _\nu |V|\Psi ^{(0)}_{\bf k}\rangle }
{E-E_\nu +\frac{i}{2}\Gamma _\nu }~.
\end{equation}
The first term on the right-hand side describes direct, or {\em potential}
\cite{Landau:77}, scattering of the positron by the ground-state molecule.
The second term describes positron capture into bound positron-molecule
states. Equation (\ref{eq:totpsi}) has the appearance of a standard
perturbation-theory formula. The energy of the system is $E=E_0+k^2/2$, where
$E_0$ is the target ground state energy. The energies of the positron-molecule
(quasi)bound states $\Phi _\nu $ in the denominator are complex,
$E_\nu -\frac{i}{2}\Gamma _\nu $, because these states are, in fact, unstable
against positron annihilation with one of the target electrons, and against
positron emission, a process inverse to positron capture. Therefore, the total
width of state $\nu $ is the sum of the annihilation and emission
(or capture) widths: $\Gamma _\nu =\Gamma _a^\nu + \Gamma _c^\nu $
\cite{note:channels}.
These states manifest as resonances in positron-molecule scattering. They may
not give a sizeable contribution to the scattering cross section, but, as
I show below, they can contribute a lot to the positron-molecule annihilation
rate.

The contribution of a particular resonant state $\nu $ to the wave function
is proportional to the corresponding capture amplitude
$\langle \Phi _\nu |V|\Psi ^{(0)}_{\bf k}\rangle $, which also determines
the capture width
\begin{equation}\label{eq:gamc}
\Gamma _c^\nu =2\pi \int |\langle \Phi _\nu |V|\Psi ^{(0)}_{\bf k}
\rangle |^2 \frac{ k d\Omega _{\bf k}}{(2\pi )^3}
=\frac{k}{\pi }|\langle \Phi _\nu |V|
\Psi ^{(0)}_{\bf k}\rangle |^2~,
\end{equation}
where the latter formula is valid for the positron $s$ wave which dominates
at low positron energies (see below). If the positron interaction with
vibrations cannot be described by perturbations Eqs. (\ref{eq:totpsi}) and
(\ref{eq:gamc}) remain valid, provided we replace the first-order
capture amplitudes $\langle \Phi _\nu |V|\Psi ^{(0)}_{\bf k}\rangle $ with
their non-perturbative values.

The annihilation width of the positron-molecule state $\Phi _\nu $ is a
product of the spin-averaged electron-positron annihilation cross section
$\overline{\sigma }_{2\gamma }=\pi r_0^2c/v$, the positron velocity $v$, and
the density factor,
\begin{eqnarray}\label{eq:gama}
\Gamma _a^\nu &=&\overline{\sigma }_{2\gamma }v\langle \Phi _\nu |
\sum _{i=1}^Z\delta ({\bf r}-{\bf r}_i)| \Phi _\nu \rangle
\nonumber \\
&=&\pi r_0^2c\int \sum _{i=1}^Z\delta ({\bf r}-{\bf r}_i)
|\Phi _\nu ({\bf r}_1,\dots ,{\bf r}_Z,{\bf r})|^2
d{\bf r}_1\dots d{\bf r}_Zd{\bf r} \\
&\equiv &\pi r_0^2c \rho ^\nu _{ep}~,\nonumber
\end{eqnarray}
where $\rho ^\nu _{ep}$ is the average positron density on the target electrons
in the $\nu$th bound state. For the ground state positronium
$\rho ^{\rm Ps}_{ep}=(8\pi a_0^3)^{-1}$. One can use this value to estimate the
annihilation width of the positron-molecule complex. The presence of many
electrons in a large molecule does not lead to an increase of the width,
because the positron is spread over a larger volume due to the normalization
condition
\begin{displaymath}
\int |\Phi _\nu ({\bf r}_1,\dots ,{\bf r}_Z,{\bf r})|^2
d{\bf r}_1\dots d{\bf r}_Zd{\bf r}=1~.
\end{displaymath}
Therefore, using the Ps estimate of the density one obtains
$\Gamma _a^\nu \sim 0.5\times 10^{-7}$ a.u. $\sim 1$ $\mu $eV, which
corresponds to the annihilation lifetime $\tau _a\sim 5\times 10^{-10}$ s.

To calculate $Z_{\rm eff}$ wave function (\ref{eq:totpsi}) is substituted
into Eq. (\ref{eq:zeffpsi}), which yields
\begin{eqnarray}\label{eq:zeff1}
Z_{\rm eff}&=&\langle \Psi _{\bf k}| \sum _{i=1}^Z
\delta ({\bf r}-{\bf r}_i)| \Psi _{\bf k}\rangle \nonumber \\
&=&\langle \Psi _{\bf k}^{(0)}| \sum _{i=1}^Z
\delta ({\bf r}-{\bf r}_i)| \Psi _{\bf k}^{(0)}\rangle +
\left\{ {{\rm interference}\atop {\rm terms}} \right\} \nonumber \\
&+&\frac{2\pi ^2}{k}\sum _{\mu \nu}\frac{A_\mu ^*\langle \Phi _\mu |
\sum _{i=1}^Z \delta ({\bf r}-{\bf r}_i)| \Phi _\nu \rangle A_\nu }
{(E-E_\mu -\frac{i}{2}\Gamma _\mu )(E-E_\nu +\frac {i}{2}\Gamma _\nu )}~,
\end{eqnarray}
where $A_\nu $ is the capture amplitude introduced as
$\Gamma _c^\nu =2\pi |A_\nu |^2$ [cf. Eq. (\ref{eq:gamc})].
The terms on the right-hand side correspond to the contributions of
direct annihilation, resonant annihilation (i.e., annihilation of the positron
captured into the positron-molecule quasibound state), and the interference
between the two.

\subsection{Direct annihilation}\label{subsec:dir}

The direct annihilation term in Eq. (\ref{eq:zeff1})
\begin{equation}\label{eq:Zdir}
Z_{\rm eff}^{\rm (dir)}=\langle
\Psi _{\bf k}^{(0)} | \sum _{i=1}^Z \delta ({\bf r}-{\bf r}_i)|
\Psi _{\bf k}^{(0)}\rangle
\end{equation}
is a smooth function of the positron energy. Let us estimate its magnitude
and find its energy dependence at small positron energies. When the
positron is outside the atomic system, $r>R_a$, the wave function
$\Psi _{\bf k}^{(0)}$ contains contributions of the incoming and scattered
positron waves
\begin{equation}\label{eq:out}
\Psi _{\bf k}^{(0)}({\bf r}_1,\dots ,{\bf r}_Z,{\bf r})=\Phi _0
({\bf r}_1,\dots ,{\bf r}_Z)\left[e^{i{\bf kr}}+f(\Omega )\frac{e^{ikr}}{r}
\right] ~,
\end{equation}
where $f(\Omega )$ is the scattering amplitude. Due to positron
repulsion from the atomic nuclei the low-energy positron does not penetrate
deep inside the atomic system. Accordingly, the positron annihilates mostly
with the outer valence electrons, where the electron and positron densities
overlap. This takes place ``on the surface'' of the atomic system, and
Eq.~(\ref{eq:out}) essentially determines the amplitude of finding the
positron there. Of course, due to short-range electron-positron correlations
the true wave function at small distances cannot be factorized similarly
to Eq. (\ref{eq:out}). The Coulomb interaction between the positron and
electron increases the probability of finding both at the same point in space,
as required by the $\delta $-function in Eq. (\ref{eq:zeffpsi}). This effect
enhances the annihilation rate \cite{Dzuba:96}. However, since small
distances and relatively
large interactions are involved, these correlations do not depend on the
momentum of the incoming positron at low energies. On the other hand,
to participate in the annihilation event the positron must first approach the
target, and this is described by Eq. (\ref{eq:out}). Unlike the
short-range correlation effects, the scattering amplitude can be very sensitive
to the positron energy. This effect is fully accounted for by
Eq. (\ref{eq:out}), and I use it to evaluate the energy dependence and
magnitude of $Z_{\rm eff}^{\rm (dir)}$.

After substitution of expression (\ref{eq:out}) into Eq. (\ref{eq:Zdir}) one
obtains
\begin{equation}\label{eq:dir1}
Z_{\rm eff}^{\rm (dir)}=\int \rho ({\bf r})
\left[e^{i{\bf kr}}+f(\Omega )\frac{e^{ikr}}{r} \right] 
\left[e^{-i{\bf kr}}+f^*(\Omega )\frac{e^{-ikr}}{r} \right] r^2drd\Omega ~,
\end{equation}
where $\rho ({\bf r})\equiv \langle \Phi _0| \sum _{i=1}^Z
\delta ({\bf r}-{\bf r}_i)| \Phi _0\rangle $ is the electron density in the
ground state of the system. The electron density drops quickly outside the
atom, and the positron density decreases rapidly inside the atom. Therefore
the integration in Eq. (\ref{eq:dir1}) should be taken over a relatively thin
shell of thickness $\delta R_a$ enclosing the atomic system. Let us
approximate the integration domain by a spherical shell of radius $r=R_a$,
where $R_a$ is the typical distance between the positron and the target during
the annihilation, comparable to the size of the atom or molecule. For small
positron momenta, $kR_a<1$, Eq. (\ref{eq:dir1}) then yields
\begin{equation}\label{eq:Zdir1}
Z_{\rm eff}^{\rm (dir)}=4\pi \rho _e\delta R_a \left( R_a^2+
\frac{\sigma _{\rm el}}{4 \pi }+2R_a{\rm Re}f_0 \right)~,
\end{equation}
where $\rho _e$ is the electron density in the annihilation range (which can
be enhanced due to short-range electron-positron correlations),
$\sigma _{\rm el}$ is the elastic cross section,
$\sigma _{\rm el}=\int |f(\Omega )|^2
d\Omega $, and $f_0$ is the spherically symmetric part of the scattering
amplitude, $f_0=(4\pi )^{-1}\int f(\Omega )d\Omega $. For positron
interaction with an atom the latter is simply equal to the $s$-wave scattering
amplitude. Its real part is expressed in terms of the $s$ wave phase shift
$\delta _0$ as ${\rm Re}f_0=\sin 2\delta _0/2k$. The $s$ wave gives a
dominant contribution to the cross section $\sigma _{\rm el}$ at low projectile
energies \cite{Landau:77}. For $k\rightarrow 0$ it is determined by the
scattering length $a$, $\sigma _{\rm el}=4\pi a^2$, as $f(\Omega )=-a$ in this
limit. A similar description is also valid for positron scattering from a
molecule at small momenta.

Note that the relation between $Z_{\rm eff}^{\rm (dir)}$ and elastic scattering
given by Eq. (\ref{eq:Zdir1}) could also be derived by matching the true
many-body wave function of the positron-target system at low energy
($E\approx 0$) with the asymptotic form (\ref{eq:out}). In this case $R_a$
will be the matching radius, and the factor before the brackets will remain
a free atomic-sized parameter. However, even in the form (\ref{eq:Zdir1}) the
electron density $\rho _e$ and the overlap $\delta R_a$ are effective
parameters, and the accurate value of the pre-factor can only be found by
comparison with numerical calculations (see Sec. \ref{subsec:illdir}).
Nevertheless, Eq. (\ref{eq:Zdir1}) is very useful for the analysis of
direct annihilation. The three terms in brackets are due to the incoming
positron plane wave, the scattered wave, and the interference term,
respectively, cf. Eqs. (\ref{eq:out}) and (\ref{eq:dir1}). Even if the cross
section $\sigma _{\rm el}$ is zero or very
small, as in the case of a Ramsauer-Townsend minimum, the annihilation rate
$Z_{\rm eff}^{\rm (dir)}$ is nonzero. Its magnitude is determined by the
effective annihilation radius $R_a$, electron density $\rho _e$ and
$\delta R_a$, which gives $Z_{\rm eff}^{\rm (dir)}\sim 1$--10, since the
quantities involved have ``normal'', atomic-size values.

Equation (\ref{eq:Zdir1}) shows that the annihilation rate for slow positrons
is greatly enhanced if the scattering cross section is large. This occurs
when the scattering length is large, because the positron-target interaction
supports a low-lying virtual $s$ level ($a<0$) or a weakly bound $s$ state
($a>0$) \cite{Landau:77}. Their energies, $\varepsilon _0=\pm 1/2a^2$,
respectively, must be much smaller than typical atomic energies,
$|\varepsilon _0|\ll 1~{\rm Ryd}$. For $|a|\gg R_a$ the scattering cross
section at low energies is much greater than the geometrical size of the
target. This effect leads to strong enhancement of $Z_{\rm eff}^{\rm (dir)}$
\cite{Goldanskii:64,Dzuba:93,Dzuba:96}. Theoretically, this gives a possibility
of infinitely large cross sections and annihilation rates at zero positron
energy, if $|a|\rightarrow \infty $. However, for nonzero momenta the $s$ wave
cross section does not exceed the unitarity limit $\sigma _{\rm el}= 4\pi /k^2$
(for the $s$ wave). This fact puts a bound on the enhancement of 
$Z_{\rm eff}^{\rm (dir)}$. For example, for thermal positrons with
$k^2/2\sim k_BT$ at room temperature ($k\sim 0.05$ a.u.) we obtain
$Z_{\rm eff}^{\rm (dir)}\sim 10^3$ from Eq. (\ref{eq:Zdir1}). Consequently,
much higher values of $Z_{\rm eff}$ cannot be produced by the direct
annihilation mechanism. A more detailed discussion of this point and
illustrations of the validity of Eq. (\ref{eq:Zdir1}) are presented in
Sec. \ref{subsec:illdir}.

\subsection{Resonant annihilation}\label{subsec:res}

Unlike the direct annihilation term, the interference and the resonant terms
on the right-hand side of Eq. (\ref{eq:zeff1}) are rapidly varying functions
of energy. The energy scale of this variation is given by the mean spacing $D$
between the resonances. If the resonances are due to vibrational excitations
of a single mode of the positron-molecule complex then $D=\omega $,
with $\omega \lesssim 0.1$ eV for a typical vibrational frequency. In a
complex molecule the positron attachment energy is sufficient for excitation
of several modes, and $D$ can be much smaller. To describe the annihilation
rates observed in experiments with non-monochromatic, e.g., thermal,
positrons, one needs to average the interference and resonance terms over an
energy interval $\Delta E$ which contains many resonances:
\begin{eqnarray}\label{eq:avE}
\frac{1}{\Delta E}\int \limits _{\Delta E} dE\left[
2\sqrt{\frac{2\pi ^2}{k}}{\rm Re}\sum _\nu
\frac{\langle \Psi _{\bf k}^{(0)}|
\sum _{i=1}^Z \delta ({\bf r}-{\bf r}_i)| \Phi _\nu \rangle A_\nu }
{E-E_\nu +\frac {i}{2}\Gamma _\nu } \right. \nonumber \\
\left. +\frac{2\pi ^2}{k}\sum _{\mu \nu}\frac{A_\mu ^*\langle \Phi _\mu |
\sum _{i=1}^Z \delta ({\bf r}-{\bf r}_i)| \Phi _\nu \rangle A_\nu }
{(E-E_\mu -\frac{i}{2}\Gamma _\mu )(E-E_\nu +\frac {i}{2}\Gamma _\nu )}
\right]
\end{eqnarray}
Upon averaging the first, interference term vanishes. In the second, resonance
term the diagonal items in the sum ($\mu =\nu $) dominate. Averaging is
then reduced to the integral over the Breit-Wigner resonant profiles. The
number of resonances within $\Delta E$ is $\Delta E/D$. Therefore, the total
annihilation rate is the sum of the direct and resonant contributions,
\begin{equation}\label{eq:result}
Z_{\rm eff}=Z_{\rm eff}^{\rm (dir)}+Z_{\rm eff}^{\rm (res)}~,
\end{equation}
with the resonant contribution given by
\begin{equation}\label{eq:Zres}
Z_{\rm eff}^{\rm (res)}=\frac{2\pi ^2}{k}
\left\langle \frac{\rho ^\nu _{ep}\Gamma _c^\nu }{D[\Gamma _a^\nu +
\Gamma _c^\nu ]}\right\rangle ~,
\end{equation}
where the angular brackets stand for averaging over the resonances, and
$\Gamma _\nu=\Gamma _a^\nu + \Gamma _c^\nu $ substituted for the total
width. Below I will show that the resonant term in Eq. (\ref{eq:result})
can be much greater than the direct one, and very high $Z_{\rm eff}$ values
can be achieved.

It is easy to see that the resonant contribution could also be derived from
standard resonant scattering theory developed originally to describe neutron
scattering via compound nucleus resonances (\cite{Landau:77}, Ch. 18) .
The maximal $s$-wave capture cross section is given by
$\sigma =\pi \lambdabar ^2\equiv \pi k^{-2}$. The true capture cross
section is smaller than $\sigma $, because the capture takes place only when
the positron energy matches the energy of the resonance. For positrons with
finite energy spread (e.g., thermal ones), the capture cross section is then
$\sigma _c\sim (\Gamma _c/D)\sigma $, where $D$ is the mean energy spacing
between the resonances. More accurately,
$\sigma _c=(2\pi \Gamma _c/D)\sigma $ \cite{Landau:77}. If we are concerned
with the annihilation process, the capture cross section must be multiplied by
the probability of annihilation, $P_a=\Gamma _a/(\Gamma _c+\Gamma _a)$,
which gives the energy-averaged resonance annihilation cross section
\begin{equation}\label{eq:sigan}
\sigma _a=\frac{2\pi ^2}{k^2}\,\frac{\Gamma _a\Gamma _c}
{D(\Gamma _c+\Gamma _a)}~,
\end{equation}
where averaging over resonances similar to that in Eq. (\ref{eq:Zres})
is assumed. By comparison with Eqs. (\ref{eq:siganzeff}) and
(\ref{eq:gama}), the resonant contribution to $Z_{\rm eff}$,
Eq. (\ref{eq:Zres}), is recovered.

The way Eq. (\ref{eq:Zres}) has been derived implies that the positrons are
captured in the $s$ wave. Otherwise, an additional factor of $(2l+1)$, where
$l$ is the positron orbital momentum, appears in the formula \cite{Landau:77}.
At low positron energies the capture widths behave as
\begin{equation}\label{eq:gamcth}
\Gamma _c \propto (kR)^{2l+1}
\end{equation}
for resonances formed by positron capture with the orbital momentum $l$
\cite{Landau:77} ($R$ is the typical radius of the target). So, the $s$ wave
capture indeed dominates in the resonant
annihilation of slow positrons. At higher energies contributions of several
lowest partial waves should be added in $Z_{\rm eff}^{\rm (res)}$.

Let us estimate the rate of resonant annihilation and compare it with the
maximal direct contribution $Z_{\rm eff}^{\rm (dir)}\sim 10^3$ for
room-temperature positrons. The typical annihilation widths for
positron-molecule (quasi)bound states are very small,
$\Gamma _a^\nu \sim 1~\mu{\rm eV}$ (see Sec. \ref{subsec:gen}). If one assumes
that the positron capture width is much greater,
\begin{equation}\label{eq:2gam}
\Gamma _c^\nu \gg \Gamma _a^\nu ,
\end{equation}
the total width $\Gamma _\nu \approx \Gamma _c^\nu $ cancels the capture
width in Eq. (\ref{eq:Zres}), and the resonant
contribution is given by
\begin{equation}\label{eq:resres}
Z_{\rm eff}^{\rm (res)}=\frac{2\pi ^2}{k}
\left\langle \frac{\rho ^\nu _{ep}}{D}\right\rangle =
\frac{2\pi ^2}{k}\rho _{ep}\rho (E_v)~.
\end{equation}
In the last equality I use the fact that electron-positron degrees of freedom
are almost unaffected by the vibrational motion of the nuclei. Hence, for a
given molecule the positron density on the target electrons $\rho _{ep}$ is
the same for different vibrational resonances. I have also introduced the
density of resonances $\rho (E_v)=D^{-1}$, where
$E_v=\varepsilon _A+\varepsilon $ is the vibrational excitation energy due to
positron-molecule binding. Equation (\ref{eq:resres}) shows that for
$\Gamma _c> 1$ $\mu $eV the contribution of the resonant mechanism is
{\em independent} of the capture width, and is determined by the density of
positron-molecule resonant states populated by positron capture. Suppose
that only a single mode with $D\sim 0.1$ eV is excited. Equation
(\ref{eq:resres}) then yields $Z_{\rm eff}^{\rm (res)}\sim 4\times 10^3$,
if I use the estimates $\rho _{ep}=\rho ^{\rm Ps} _{ep}$, and $k=0.05$ for
room-temperature positrons.

The resonance spacing $D$ cannot be smaller than the widths of the resonances,
which are limited by the annihilation width $\Gamma _a$. Thus, one can obtain
an upper estimate of the resonant annihilation rate from Eq. (\ref{eq:Zres})
by putting $\Gamma _c\approx \Gamma _a\sim 0.5\times 10^{-7}$ a.u., and 
$D\sim 2\pi \Gamma _c$, which gives the maximal possible capture cross section
$\sigma $. These estimates yield $Z_{\rm eff}^{\rm (res)}\sim 5\times 10^7$
at room temperature (cf. $Z_{\rm eff}=7.5\times 10^6$ for C$_{12}$H$_{22}$0$_4$
\cite{Leventhal:90}). This theoretical maximum of $Z_{\rm eff}^{\rm (res)}$ 
corresponds to the unitarity limit of the $s$ wave capture cross section.
However, this estimate of $Z_{\rm eff}$ is not trivial. The resonance
mechanism shows that such large cross sections can be achieved for the
annihilation process, in spite of the fact that it is suppressed by the
relativistic factor $\pi r_0^2c=\pi /c^3 \sim 10^{-6}$, in atomic units [see
Eq. (\ref{eq:siganzeff})].

Equation (\ref{eq:resres}) predicts unusual low-energy threshold
behaviour $Z_{\rm eff}^{\rm (res)}\propto 1/k\propto 1/\sqrt{T}$ (the latter
for thermal positrons). In a standard situation the cross section of an
inelastic process involving a slow projectile in the initial state
behaves as $\sigma \propto 1/k$. This dependence is characteristic of the
$s$ wave scattering, which dominates at low projectile energies, and is
valid in the absence of long-range forces between the target and the
projectile. It is known as the ``$1/v$'' law, and its examples are numerous:
from the $(n,\gamma )$ nuclear reaction to dissociative electron attachment to
molecules, where it is observed below 1 meV \cite{Klar:92}.
Therefore, one would expect the positron annihilation cross section to
behave as $\sigma _a\propto 1/k$. Accordingly, $Z_{\rm eff}$, which is
proportional the annihilation rate, is expected to be constant
at low positron energies.

The anomalous threshold dependence of Eq. (\ref{eq:resres}) clearly
contradicts this general statement. This ``puzzle'' is easily resolved if we
recall condition (\ref{eq:2gam}) that has lead to Eq. (\ref{eq:resres}). For
very low positron momenta the $s$-wave capture width behaves as
$\Gamma _c\propto kR$, so that (\ref{eq:2gam}) is clearly violated, and the
resonant contribution in Eq. (\ref{eq:result}) becomes constant as
$k\rightarrow 0$. However, at higher positron energies the $1/k$ behaviour of
$Z_{\rm eff}$ may be observed. This dependence corresponds to the
$1/\varepsilon $ drop of the cross section which is reported in some electron
attachment experiments (see, e.g., \cite{Kiendler:96}).

The fact that positron-molecule resonances give a large contribution to the
annihilation rate, as compared to the direct annihilation, does not mean that
they also contribute much to the elastic scattering cross section. In analogy
with Eq. (\ref{eq:sigan}), the resonant contribution to the elastic scattering
is given by
\begin{equation}\label{eq:sigelres}
\sigma _{\rm el}^{\rm (res)}=\frac{2\pi ^2}{k^2}\,\frac{\Gamma _c^2}
{D(\Gamma _c+\Gamma _a)}~,
\end{equation}
and for $\Gamma _c\ll D$ it is much smaller than the direct, or potential,
scattering cross section.

%***************************************************************************
\section{Illustrations and comparison with experiment}\label{sec:ill}

\subsection{Effect of virtual or weakly bound states on direct
annihilation}\label{subsec:illdir}

If low-energy positron scattering is dominated by the presence of a virtual
or weakly bound state at $\varepsilon _0=\pm \kappa ^2/2$, the corresponding
cross section has the form (for scattering by a short-range potential
\cite{Landau:77}) 
\begin{equation}\label{eq:csvirt}
\sigma _{\rm el}=\frac{4\pi }{\kappa ^2 +k^2}~,
\end{equation}
where $\kappa =a^{-1}$.
According to Eq. (\ref{eq:Zdir1}) a similar maximum should appear in the
momentum dependence of the annihilation rate. Its magnitude at $k= 0$ can be
arbitrarily large if $\kappa \rightarrow 0 $ ($|a|\rightarrow \infty$), which
corresponds to a level at zero energy. However, for nonzero momenta the
maximal cross section is finite, $\sigma _{\rm el}\sim 4\pi /k^2$, which
corresponds to the unitarity limit for the $s$-wave cross section.

Real atomic and molecular targets have nonzero electric dipole
polarizabilities $\alpha $, which give rise to the long-range polarization
potential $-\alpha /2r^4$ for the positron. Its effect is taken into account
by the modified effective-range formula for the $s$-wave phase shift
\cite{OMalley:61},
\begin{eqnarray}\label{eq:MER}
\tan \delta _0 = -ak\left[1-\frac{\pi \alpha k}{3a}-\frac{4\alpha k^2}{3}
\ln \left( \frac{C}{4}\sqrt{\alpha }k \right)\right]^{-1}~, \\
\sigma _{\rm el}=\frac{4\pi a^2}{ \left[ 1-(\pi \alpha k/3a)-(4\alpha k^2/3)
\ln \left( \frac{C}{4}\sqrt{\alpha }k \right) \right]^2+a^2k^2 }~,
\label{eq:MERsig}
\end{eqnarray}
the latter formula being valid when the scattering length is large and the
$s$-wave scattering dominates at small $k$. In equations (\ref{eq:MER})
and (\ref{eq:MERsig}) $C$ is a dimensionless positive constant. Note that for
$\alpha =0$, Eq. (\ref{eq:csvirt}) is immediately recovered. The polarization
potential modifies the behaviour of the cross section at low energies. For
example, it leads to a more rapid decrease of the cross section for $a<0$,
$\sigma _{\rm el}=4\pi a^2 [1+2\pi \alpha k/3a +O(k^2\ln k)]$. However, this
does not change the estimates of the maximal values of $Z_{\rm eff}$ that
could be produced in direct annihilation.

\begin{figure}[t]
\vspace{-100pt}
\epsfxsize=15.000cm
\centering\leavevmode\epsfbox{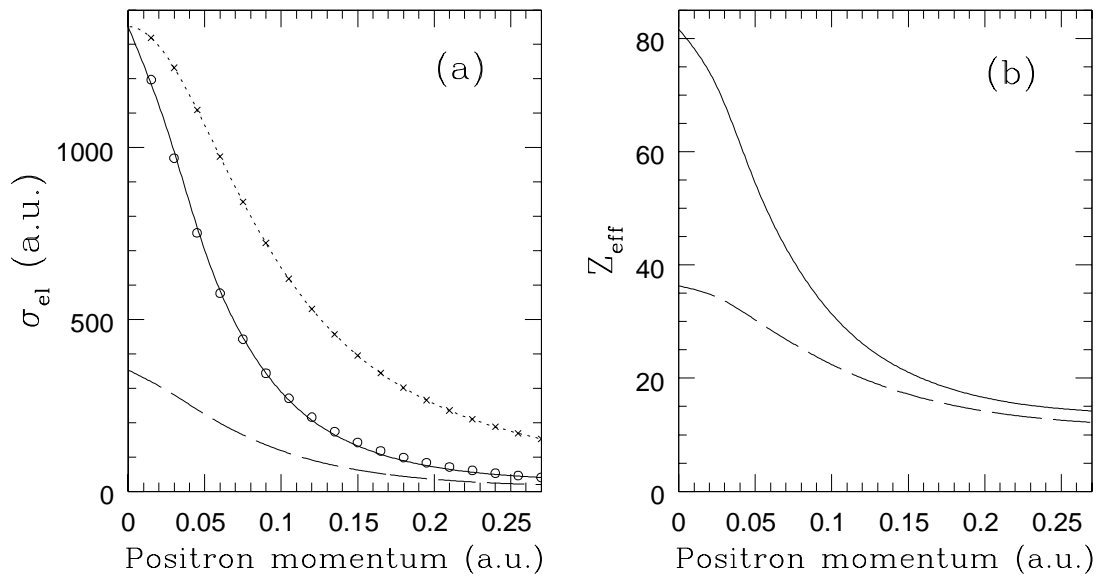}
\vspace{-90pt}
\caption{Elastic scattering cross section $\sigma _{\rm el}$ (a) and
annihilation rates $Z_{\rm eff}$ (b) for Ar (dashed curves) and Kr
(solid curves), as calculated in Ref. \protect \cite{McEachran:79}. Also
shown in (a) are the analytical approximations of $\sigma _{\rm el}$ for Kr
by the short-range potential formula (\protect \ref{eq:csvirt}) (dotted line
with crosses), and the modified effective range formula
(\protect \ref{eq:MERsig}), which accounts for the dipole
polarization of the target (open circles). Here I have used the calculated
value of $a=-10.4$ a.u., experimental dipole polarizability $\alpha =16.74$
a.u. \protect \cite{Radtsig:86}, and $C=0.4$ obtained from the $s$-wave phase
shift of Ref. \protect \cite{McEachran:79}. Note that the modified effective
range formula (open circles) gives an accurate description of the cross
section shown by the solid curve.
}
\label{fig:ZandCS}
\end{figure}

To illustrate the relation between direct annihilation and elastic scattering,
and the enhancement of both due to the presence of a low-lying virtual level,
let us compare the behaviour of $Z_{\rm eff}$ and $\sigma _{\rm el}$ for
Ar and Kr. The results shown in Fig. \ref{fig:ZandCS} were obtained within
the polarized-orbital method \cite{McEachran:79}, which takes into account the
polarization of the target by the positron. These calculations yield large
negative values of the scattering length for Ar, Kr and Xe (see Table
\ref{tab:aR}), indicating the presence of positron-atom virtual levels
formed due to strong positron-atom attraction. The increase of $|a|$
correlates with the
increase of the dipole polarizability in these atoms. Similar values of $a$
have been obtained in the many-body theory calculations of Dzuba {\em et al.}
\cite{Dzuba:96}. Figure \ref{fig:ZandCS} shows that both $\sigma _{\rm el}$ and
$Z_{\rm eff}$ are enhanced at low momenta due to the presence of the virtual
$s$ levels. This effect is stronger for Kr, which has a greater absolute value
of the positron scattering length. As illustrated by Fig. \ref{fig:ZandCS}a
for Kr, Eq. (\ref{eq:MERsig}) provides a good description of the cross section
at  small $k$. The visible difference between $Z_{\rm eff}$ and
$\sigma _{\rm el}$ in Fig. \ref{fig:ZandCS} is due to the background given by
the energy-independent term $R_a^2$ in Eq. (\ref{eq:Zdir1}).

\begin{figure}[h]
%\vspace{-100pt}
\epsfxsize=14.000cm
\centering\leavevmode\epsfbox{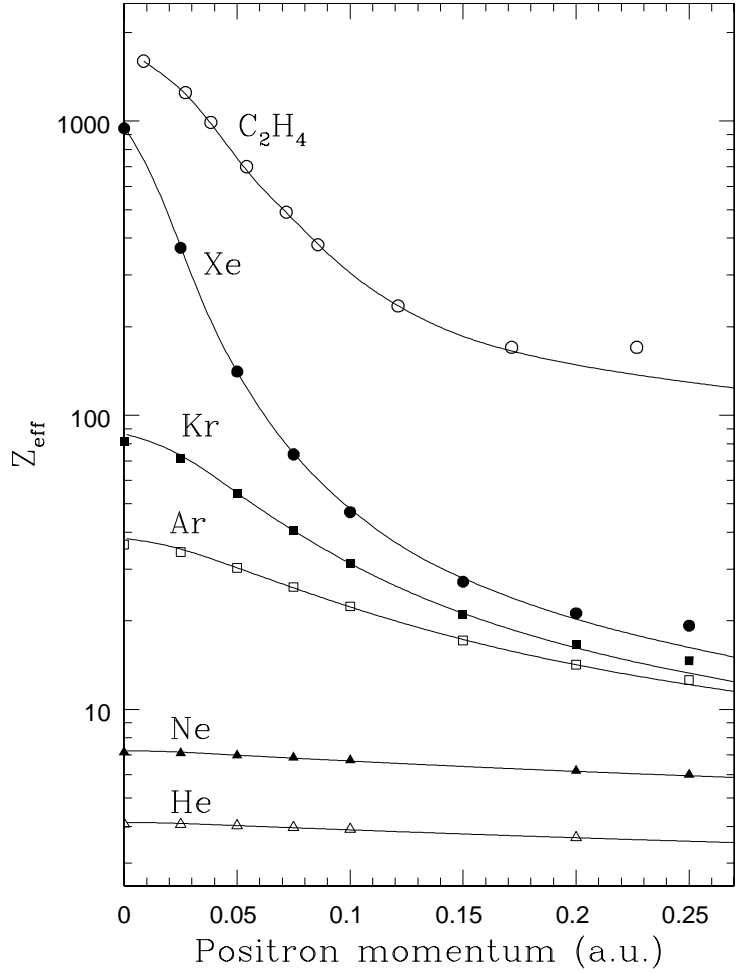}
%\vspace{-90pt}
\caption{Relation between $Z_{\rm eff}$ due to direct annihilation
and the elastic scattering cross section. Calculated $Z_{\rm eff}$ values for
He (open triangles), Ne (solid triangles), Ar (open squares),
Kr (solid squares), and Xe (solid circles) \protect \cite{McEachran:79}, and
C$_2$H$_4$ (open circles) \protect \cite{daSilva:96} are compared with the
predictions of Eq. (\protect \ref{eq:Zdir1}), shown by solid curves. In the
latter I have used the scattering cross sections and amplitudes calculated
in the same theoretical papers, and considered $R_a$ and the pre-factor
$4\pi \rho _e\delta R_a$ as fitting parameters.}
\label{fig:Zdir}
\end{figure}

Figure \ref{fig:Zdir} provides a direct comparison between $Z_{\rm eff}$
and the right-hand side of Eq. (\ref{eq:Zdir1}), and shows that this
relation is valid at low positron energies. The comparison is based on the
polarized-orbital method results for the noble-gas atoms \cite{McEachran:79},
and the values of $Z_{\rm eff}$ and $\sigma _{\rm el}$ obtained for the
ethylene molecule (C$_2$H$_4$) by the Schwinger multichannel method
\cite{daSilva:96}. In this comparison I have considered $R_a$ and the
pre-factor $4\pi \rho _e\delta R_a$ in Eq. (\ref{eq:Zdir1}) as fitting
parameters. Their values are listed in Table \ref{tab:aR} together with the
values of $a$ obtained in those calculations. Note that the theoretical
results used to produce this plot are not necessarily ``exact'' or accurate
(although, experimental data confirm that they are reasonable
\cite{Iwata:99,Kurz:96}). It follows from the derivation that
Eq. (\ref{eq:Zdir1}) holds for any calculation, as long as the same wave
function is used in the scattering and annihilation calculations
\cite{hydrogen}.

In agreement with the estimates made in Sec. \ref{subsec:dir},
Fig. \ref{fig:Zdir} shows that direct annihilation is indeed strongly enhanced
by the presence of low-lying virtual levels. Nevertheless, even for targets
with very large scattering lengths, such as Xe or C$_2$H$_4$, the annihilation
rates do not exceed $Z_{\rm eff}\sim 10^3$ for room-temperature positron
momenta (0.05 a.u.).

Direct annihilation is the only annihilation mechanism for atoms and molecules
which do not form bound states with positrons. It will also dominate for small
molecules which do form a weakly bound state with the positron, but whose
vibrational frequencies are high. In this case the energy
$\varepsilon _A+\varepsilon $ is simply insufficient for the excitation of
the resonant quasibound states at low impact positron energies $\varepsilon $.

For large molecules the difference between the resonant and direct mechanisms
is probably most obvious when one compares the experimental values of
$Z_{\rm eff}$ for alkanes and perfluorinated alkanes shown in
Fig.~\ref{fig:Zeffmol}. The large annihilation rates of the
alkane molecules with more than two carbon atoms cannot be explained by direct
annihilation. They also display a very rapid increase with the size of the
molecule, which is typical of resonant annihilation. On the other hand, the
$Z_{\rm eff}$ values of the perfluorinated alkanes remain comparatively
small, in spite of their softer vibrational spectra. Thus, one is lead
to conclude that the resonant mechanism is switched off for them. The latter
is explained by the very weak attraction between the positron and fluorine
atoms \cite{Iwata:99}, insufficient to provide positron-molecule binding.

Let us examine the effect of fluorination on $Z_{\rm eff}$ for the lightest
molecule of the series, methane. The experimental data at room temperature
are: $Z_{\rm eff}=158.5$, 715, 411, 127, and 38, for CH$_4$, CH$_3$F,
CH$_2$F$_2$, CHF$_3$, and CF$_4$, respectively  (data from
\cite{Iwata:99,Iwata:PhD} normalized to the given value for methane).
These values are small enough to be accounted for by the direct mechanism.
Within its framework the increase and subsequent drop of $Z_{\rm eff}$ could
be explained by the existence of a loosely bound state for the positron on
methane, which turns into a virtual level as the number of substitute fluorine
atoms increases \cite{note:pol}. In terms of $\kappa $ parameter this would
mean that $\kappa $ is small and positive for CH$_4$, and then goes through
zero, and becomes negative upon fluorination. Accordingly, both the cross
section and the annihilation rate peak for the molecule with the smallest
absolute value of $\kappa $, namely CH$_3$F. This picture is considered in
Ref. \cite{Iwata:99} in more detail using the zero-range potential model for
positron-molecule interaction. 

\begin{figure}[h]
\epsfxsize=14.0cm
\centering\leavevmode\epsfbox{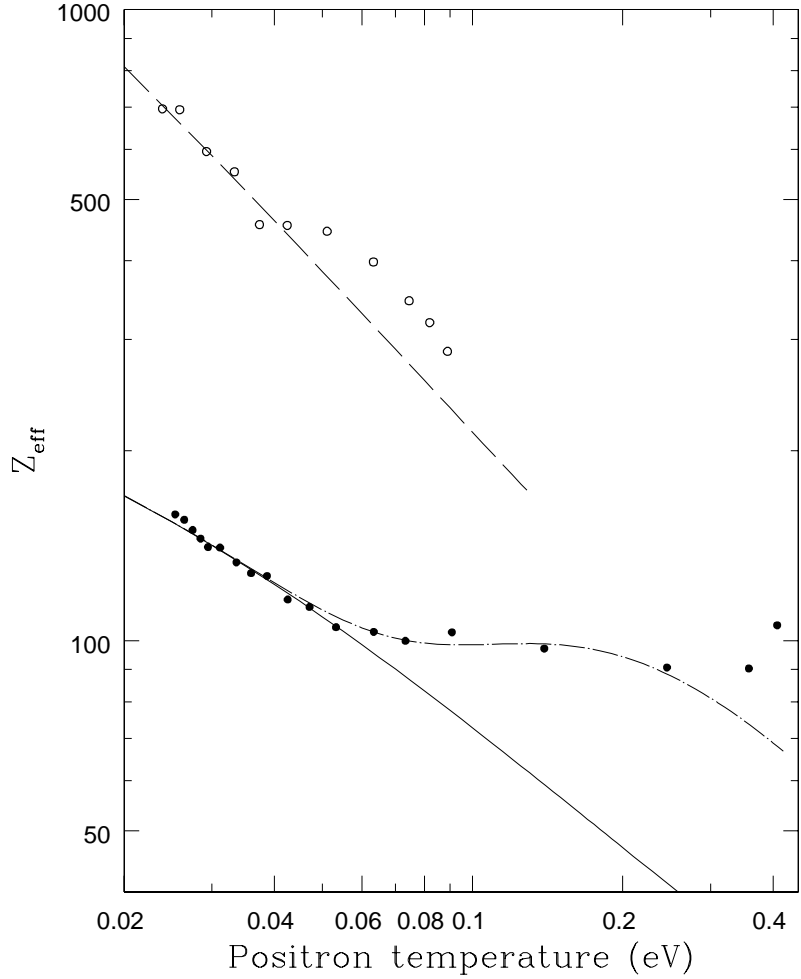}
\caption{Annihilation rates for methane and fluoromethane. Experimental data
for CH$_4$ (solid circles) and CH$_3$F (open circles) \protect \cite{Iwata:99}
have been normalized to $Z_{\rm eff}=158.5$ for methane
at room temperature. Thermal-averaged direct annihilation fits obtained from
Eqs. (\protect \ref{eq:Zdir1}) and (\protect \ref{eq:MERsig}) using
$4\pi \rho _e\delta R_a=1$, $R_a=4$, $C=1$, $\alpha =17.6$ a.u., are shown
for CH$_4$ ($\kappa =0.045$, solid curve), and CH$_3$F ($\kappa =0.01$,
dashed curve). Also shown for methane is the sum of the direct
contribution and that of the first vibrational $A_1$ resonance at
$\varepsilon _\nu = 0.33$ eV, obtained using
$\rho _{ep}^\nu =0.25\rho _{ep}^{\rm Ps}$, Eq. (\ref{eq:singres})
(chain curve).
}
\label{fig:ch4ch3f}
\end{figure}

Besides having a larger value of $Z_{\rm eff}$, the molecule with a smaller
$|\kappa |$ (i.e., larger $|a|$) should have a more rapid dependence of
the annihilation rate on the positron energy, cf. Figure \ref{fig:Zdir}.
If the experiment is done with thermal positrons this should manifest in a
stronger temperature dependence of the Maxwellian average of $Z_{\rm eff}(k)$
\begin{equation}\label{eq:ZT}
\overline{Z}_{\rm eff}(T)=\int _0^\infty \frac{e^{-k^2/2k_BT}}
{(2\pi k_BT)^{3/2}} Z_{\rm eff}(k) 4\pi k^2dk
\end{equation}
on the positron temperature $T$. The overbar is usually omitted, as it is
clear from the context whether one is dealing with $Z_{\rm eff}(k)$ at a
specific positron momentum, or with a thermal average
$\overline{Z}_{\rm eff}(T)$. The temperature dependences of the annihilation
rates for methane and fluoromethane measured in Ref. \cite{Iwata:99} are shown
in Fig. \ref{fig:ch4ch3f}. Also shown are low-temperature theoretical fits
obtained using Eqs. (\ref{eq:Zdir1}), (\ref{eq:MERsig}) and (\ref{eq:ZT}).
Their parameters are given in the caption.

The dipole polarizability of CH$_3$F $\alpha =16.1$ a.u. is close to that of
methane, $\alpha =17.6$ a.u., and I use the latter for both molecules. The
constant $C$ appears in Eqs. (\ref{eq:MER}) and (\ref{eq:MERsig}) under the
logarithm, and the result is not very sensitive to it, so $C=1$ has been
chosen. The value of the characteristic radius $R_a=4$ a.u. is similar to
those for noble gas atoms and ethylene (table \ref{tab:aR}), and the pre-factor
$4\pi \rho _e \delta R_a=1$ is between those for noble gas atoms and
C$_2$H$_4$. Of course, the number of independent parameters ($a$, $C$, $R_a$
and $4\pi \rho _e \delta R_a$) is too large to enable their unique
determination from the experimental data. However, the fits clearly demonstrate
that very different $Z_{\rm eff}(T)$ curves can be obtained {\em only} due to
different $\kappa $ values ($\kappa =0.045$ and 0.01, for CH$_4$ and CH$_3$F,
respectively. These values imply that both molecules have bound states with
the positron. The binding energy for CH$_4$ is $\varepsilon _A=\kappa ^2/2=
1.0 \times 10^{-3}$ a.u.$=0.028$ eV, and the binding
energy corresponding to $\kappa =0.01$ is just 1 meV. There is a large
uncertainty in the latter value, because measurements performed
at and above room temperature, $T=0.0253$ eV, are not really sensitive to such
small $\kappa $. This can be seen, e.g., from Eq. (\ref{eq:csvirt}), which
becomes $\kappa $-independent for $\kappa \ll k$. Zero-range model
calculations presented in Ref. \cite{Iwata:99} show that the last three
members of the fluoromethane sequence have negative $\kappa $, corresponding
to virtual levels with increasing energies. This causes the decrease of their
$Z_{\rm eff}$ values.

As seen in Fig. \ref{fig:ch4ch3f}, equation (\ref{eq:Zdir1}) for the direct
annihilation combined with the modified effective range formula
(\ref{eq:MERsig}) works well in the low-energy part of the graph. However,
the data for methane clearly show an abrupt departure from this law at higher
$T$, and the formation of some kind of a plateau in $Z_{\rm eff}(T)$. In
principle, one could think that this is due to contributions of higher partial
waves, not included in $\sigma _{\rm el}$, Eq. (\ref{eq:MERsig}). However,
their contribution has been included via the $R_a$ term of
Eq. (\ref{eq:Zdir1}).
Also, the contributions of higher partial waves to $Z_{\rm eff}$ emerge as
$\varepsilon ^l$, which is a manifestation of the Wigner threshold law
\cite{Landau:77}. For thermally averaged rates this corresponds to $T^l$.
Thus, it cannot be responsible for this sudden feature.

On the other hand, if the methane molecule forms a bound state with the
positron the system can also have vibrationally excited positron-molecule
resonant states. The positron bound state on CH$_4$ must belong to the $A_1$
symmetry type of the molecule. Since the positron $s$ wave dominates at low
energies, its capture into the $A_1$ state can result in the excitation of
$A_1$ vibrational modes of the molecule. The frequency of this mode for
methane is $\omega =2916$ cm$^{-1}=0.361$ eV. Assuming that the
positron binding does not change this frequency much, the lowest vibrationally
excited positron-molecule resonance will occur at $\varepsilon =\omega -
\varepsilon _A\approx 0.33$ eV.

It is easy to estimate the contribution of a single narrow vibrational
resonance located at positron energy $\varepsilon _\nu $ to the thermally
averaged $Z_{\rm eff}$ \cite{note:res},
\begin{equation}\label{eq:singres}
\Delta Z_{\rm eff}(T)= \frac{8\pi ^3\rho _{ep}^\nu \Gamma _c^\nu }
{\Gamma _a^\nu +\Gamma _c^\nu }\,
\frac{e^{-\varepsilon _\nu /k_BT}}{(2\pi k_BT)^{3/2}}\simeq
8\pi ^3\rho _{ep}^\nu \frac{e^{-\varepsilon _\nu /k_BT}}{(2\pi k_BT)^{3/2}}~,
\end{equation}
the latter formula valid for $\Gamma _c^\nu \gg \Gamma _a^\nu $, which
implies that the resonance has a capture width greater than 1 $\mu $eV.
Figure \ref{fig:ch4ch3f} shows the effect of the lowest vibrational
$A_1$ resonance at $\varepsilon _\nu =0.33 $ eV on $Z_{\rm eff}$ for methane
(chain curve). Its onset is indeed quite
rapid, due to the exponent in Eq. (\ref{eq:singres}), which makes
$\Delta Z_{\rm eff}(T)$ very small for $k_BT<\varepsilon _\nu $. To fit
the experimental data the density $\rho _{ep}^\nu $ is chosen to be 25\% of
$\rho _{ep}^{\rm Ps}$. One could expect that for a weakly bound state
($\varepsilon _A=0.028$ eV), where the positron spends most of its time
outside the molecule, its density on the electrons is reduced below
that of Ps (binding energy 6.8 eV) \cite{Ryzhikh:98}.

\subsection{Resonant annihilation: molecular
vibrations and temperature dependence}\label{subsec:illres}

\subsubsection{Vibrations.}

Equation (\ref{eq:resres}) derived in Sec. \ref{subsec:res} shows that the
annihilation rate due to positron capture into resonances is
determined by the level density of these quasibound vibrationally
excited states of the positron-molecule complex. This density depends
on the excitation energy available, as defined by the positron kinetic energy
and positron affinity, $E_v=\varepsilon _A+\varepsilon $, and also on the
structure of the molecular vibrational spectrum. Suppose that the molecule
possesses a particular symmetry, which is true for most of the
molecules where positron annihilation has been studied so far \cite{Iwata:95}.
The electronic ground state wave function of the molecule is usually
nondegenerate and invariant under all symmetry transformations. Let us call
this symmetry type $A$. Depending on the actual symmetry of the molecule this
can be $A_1$, $A_g$, or $A_{1g}$. If the positron can be bound by such
molecule, the electron-positron part of the wave function of the
positron-molecule complex will also be fully symmetric, i.e., of the $A$
symmetry type.

Consider now the capture of a continuous spectrum positron into the bound
positron-molecule state. At low positron energies this process is dominated
by the incident positron $s$ wave, higher partial waves being suppressed
as $(kR)^{2l}$, compared to the $s$ wave [cf.
Eq. (\ref{eq:gamcth})]. As a result, the electron-positron part of the wave
function of the initial (molecule and the $s$-wave positron) and final
(bound positron-molecule complex) states of the capture process are
characterized by the same full molecular symmetry $A$. This imposes a
selection rule on the nuclear vibrations which can be excited during the
capture process. They must also belong to the $A$ symmetry type.

Therefore, the selection rule limits the spectrum of possible vibrationally
excited resonances which could in principle be formed. It allows arbitrary
excitations and combinations of the $A$ modes. It also allows overtones and
combinations of the other symmetry types, provided such excitations contain
the $A$ symmetry type, i.e., the (symmetric) product of the symmetry types
involved contains $A$ among its irreducible representations \cite{Landau:77}.
This does not mean
that all such vibrations will contribute to the density factor $\rho (E_v)$
in Eq. (\ref{eq:resres}) for $Z_{\rm eff}$. Some of them may have extremely
weak coupling to the electron-positron degrees of freedom, with capture
widths much smaller than 1 $\mu $eV. In this case they will be effectively
decoupled from the positron capture channel, and hence, will not contribute to
$Z_{\rm eff}$. Of course, this can only be found out by doing detailed
calculations for specific molecules.

Nevertheless, it is instructive to compare Eq. (\ref{eq:resres}) with
experimental data. This comparison enables one to extract the effective
mean spacing $D$ between the positron-molecule resonances. For experiments
with thermal positrons Eq. (\ref{eq:resres}) must be averaged over the
Maxwellian positron momenta distribution,
\begin{equation}\label{eq:ZresT}
Z_{\rm eff}^{\rm (res)}=\frac{2\pi ^2\rho _{ep}}{D}\left\langle \frac{1}{k}
\right\rangle _T=\frac{2\pi ^2\rho _{ep}}{D}
\left( \frac{2}{\pi k_BT}\right) ^{1/2} .
\end{equation}
Let us use the Ps value, $\rho _{ep}=1/8\pi $, to estimate the
electron-positron density, and apply Eq. (\ref{eq:ZresT}) to simple symmetric
molecules with $Z_{\rm eff}\gtrsim 10^4$, where resonant annihilation must be
the dominant mechanism. The effective spacings
$D=4.51\times 10^{6}/Z_{\rm eff}$ (in cm$^{-1}$) obtained from the
experimental $Z_{\rm eff}$ values measured with room-temperature positrons
\cite{Iwata:95} are listed in Table \ref{tab:vibfreq}. They are compared
with the low frequency vibrational modes of the $A$ symmetry type of these
molecules taken from Ref. \cite{Sverdlov:74}. As discussed above, vibrations
of the $A$ symmetry type also occur in overtones and combinations of other
modes. However, their frequencies scale with the size and chemical composition
of the molecule in a way similar to the $A$ modes, and the $A$ mode frequencies
listed in the table are representative of the lower vibrational modes on the
whole.

For molecules with moderate $Z_{\rm eff}$ at the top of the table, such as
CCl$_4$, the effective resonance spacing $D$ is comparable to the frequencies
of single modes. With the increase of the size of the molecule (alkanes), or
masses of the constituents (e.g., CBr$_4$), the vibrational modes are softened,
and the number of low-frequency modes increases. At the same time one can
expect that the positron binding energy increases for these molecules. These
effects, and especially the increase of the number of modes, facilitate
multimode excitations, whose density is much greater that the level density of
the individual modes. Accordingly, we see that $D$ becomes much smaller that
the frequencies of the individual modes at the bottom of the table.

In the simplest model this effect can be estimated as follows. Suppose the
vibrational modes in question are characterized by some typical frequency
$\omega $, and the molecule has $N_v$ such modes. Suppose, the positron
binding energy is $\varepsilon _A=n\omega $, where $n$ is the number of
vibrational quanta excited due to positron binding. If we neglect the small
kinetic energy of the positron, $E_v\approx \varepsilon _A$, the total number
of various vibrational excitations at energy $E_v$ is given by
$(N_v+n-1)!/[n!(N_v-1)!]$ (number of ways to distribute $n$ vibrational quanta
among $N_v$ modes). For large molecules $\varepsilon _A$ remains finite,
whereas $N_v$ increases linearly with the size of the molecule, the total
number of vibrational modes being $3N-6$, where $N$ is the number of atoms.
Therefore, the number of vibrational excitations available, and the density
of the resonant vibrational spectrum, increase as $(N_v)^n\propto N^{n}$.
Such rapid increase is indeed observed for alkanes and aromatic hydrocarbons,
see Fig. \ref{fig:Zeffmol}. The effective number of vibrational modes excited
in the capture process, $n=6.1$ and 8.2, respectively, is compatible with
the positron binding energy of few tens of an electron Volt. For example,
if I use the lowest $A_g$ mode frequency of hexane (Table \ref{tab:vibfreq}),
the positron affinity is $\varepsilon _A\sim 6\omega \approx 0.25$ eV.
This number looks reasonable, compared with positron binding
energies on single atoms, e.g., $\varepsilon _A=0.08$, 0.15, and 0.38,
for Be, Cu and Mg, respectively \cite{Ryzhikh:97,Ryzhikh:98}.

Apart from the rapid growth, $Z_{\rm eff}$ for alkanes shows clear signs of
saturation, when the number of carbon atoms becomes greater than 8 or 10.
Apparently, this takes place well before the unitarity limit derived in Sec.
\ref{subsec:res} is reached. This behaviour can be understood if we recall that
Eq. (\ref{eq:resres}) is valid only when the capture width $\Gamma _c$ is
greater than the annihilation width $\Gamma _a$. With the increase of the
number of vibrational modes their coupling to the electron-positron degrees
of freedom decreases. This coupling is represented by $\Gamma _c$,
and for small capture widths, $\Gamma _c<\Gamma _a$,
$Z_{\rm eff}^{\rm (res)}$ from Eq. (\ref{eq:Zres}) is estimated as
\begin{equation}\label{eq:ZsmG}
Z_{\rm eff}^{\rm (res)}\simeq \frac{2\pi ^2}{k}
\left \langle \frac{\rho _{ep}^\nu \Gamma _c^\nu }{D\Gamma _a^\nu }
\right \rangle 
=\frac{2\pi c^3}{k}\left\langle \frac{\Gamma _c^\nu }{D}\right \rangle ~,
\end{equation}
where Eq. (\ref{eq:gama}) is used together with $r_0=c^{-2}$, in atomic units.
The decrease of $\Gamma _c$ is a simple consequence of sum rules, because the
total strength of
positron coupling is distributed among larger number of possible vibrational
excitations. In this regime $\Gamma _c$ is proportional to $D$,
and the increase of $Z_{\rm eff}^{\rm (res)}$ related to the increase of the
density of vibrational excitation spectrum stops. The relation
$\Gamma _c \propto D$ which characterizes this regime is well known in neutron
capture into compound resonances \cite{Bohr:69}. It takes place in complex
atomic spectra, e.g., in rare-earths, where the oscillator strengths are
distributed among very large numbers of transitions \cite{Sobelman:92}.
It also emerges in the unimolecular reaction treatment of dissociative
electron attachment \cite{Christo:84}, where it is responsible for very large
lifetimes (i.e., small state widths) of transient molecular anions.

\subsubsection{Dependence on the positron energy or temperature.}

\begin{figure}[h]
\epsfxsize=10.0cm
\centering\leavevmode\epsfbox{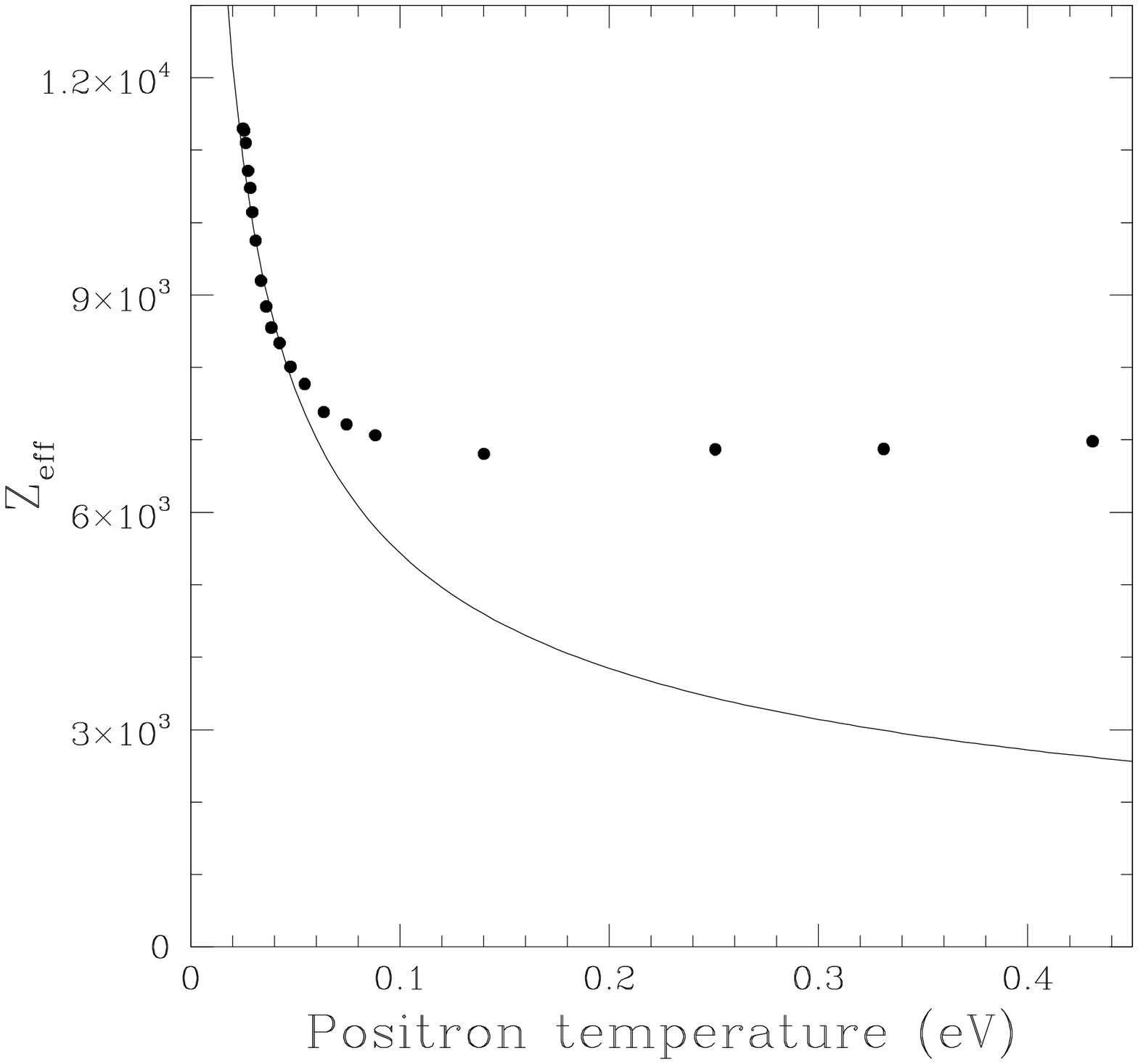}
\caption{Dependence of $Z_{\rm eff}$ on positron temperature for butane,
C$_4$H$_{10}$. Solid circles, experimental data \protect \cite{Iwata:99},
normalized at room temperature to $Z_{\rm eff}=11300$ \protect \cite{Iwata:95}.
Solid curve is the $1/\sqrt{T}$ dependence, Eq. (\protect \ref{eq:ZresT}),
with $\rho _{ep}=\rho _{ep}^{\rm Ps}$, and effective resonance spacing
$D=1.90\times 10^{-3}\,{\rm a.u.}=417$ cm$^{-1}$.}
\label{fig:c4h10}
\end{figure}

Let us now look at the energy dependence of the resonant annihilation rate.
At very small positron energies $Z_{\rm eff}^{\rm (res)}$ must be constant
(see discussion at the end of Sec. \ref{subsec:res}). However, as soon as the
$s$-wave capture width becomes greater that 1 $\mu $eV, the corresponding
annihilation rate shows a $1/k\sim \varepsilon ^{-1/2}$ dependence on positron
energy, as predicted by Eq. (\ref{eq:resres}). For a thermally averaged
rate this is described by Eq. (\ref{eq:ZresT}). Figure \ref{fig:c4h10} presents
a comparison between the $1/\sqrt{T}$ law and the experimental temperature
dependence of $Z_{\rm eff}$ for C$_4$H$_{10}$ \cite{Iwata:99}. This molecule
has $Z_{\rm eff}\sim 10^4$. Within the present theoretical framework this
large value must be due to the resonant annihilation process.

The theory and experiment agree well at low temperatures. One may
notice that the measured $Z_{\rm eff}$ show a slightly steeper rise towards
small $T$. However, the difference is not large, both in relative and absolute
terms. It could be explained by a direct contribution
$Z_{\rm eff}^{\rm (dir)}$ in Eq. (\ref{eq:result}), which peaks sharply
at small energies, if the positron-molecule scattering length is large
(see Sec. \ref{subsec:illdir}). In spite of the dominance of the resonant
contribution, $Z_{\rm eff}^{\rm (res)}\sim 10^4$ for butane, the addition of
$Z_{\rm eff}^{\rm (dir)}\sim 10^3$ at small positron energies would still
be noticeable.

A more pronounced feature of the experimental data, which is not accounted for
by Eq. (\ref{eq:ZresT}), is the plateau observed at higher temperatures,
$T>0.05$ eV, where $Z_{\rm eff}$ goes well above the $1/\sqrt{T}$ curve.
To find its possible origins let us first take a closer look at
Eq. (\ref{eq:ZresT}) and its predecessor, Eq. (\ref{eq:resres}). For small
impact positron energies $\varepsilon $ the vibrational excitation energy
is given by $E_v\approx \varepsilon _A$. Accordingly, the resonance density
$\rho (E_v)$ in Eq. (\ref{eq:resres}), and the mean spacing $D$ in
Eq. (\ref{eq:ZresT}) are approximately constant. As the positron energy, or
temperature, increase, the resonance density factor should also increase,
since $\rho (E_v)$ is a strong function of the excitation energy for multimode
vibrational spectra. Therefore, the decrease of $Z_{\rm eff}^{\rm (res)}$
should be slower than $1/k$, or $1/\sqrt{T}$. Moreover, the density factor
may even produce a rise in the energy dependence of $Z_{\rm eff}^{\rm (res)}$.
Besides this, contributions of higher positron partial waves which
emerge as $T$, $T^2$, etc., at small $T$, may also contribute to
$Z_{\rm eff}^{\rm (res)}$ in the plateau region. It might even seem that these
effects could lead to a rapid increase of $Z_{\rm eff}^{\rm (res)}$ with
positron energy.

However, there is an effect that suppresses the increase of resonant
annihilation. Throughout the paper I have assumed that the positron-molecule
resonances have only two decay channels, annihilation and detachment, the
latter being the reverse of positron capture. When the positron energy
rises above the threshold of molecular vibrational excitations, the resonances
can also decay into the `positron $+$ vibrationally excited molecule' channels.
In this situation the total width of a resonance will be
given by $\Gamma _\nu =\Gamma _a^\nu +\Gamma _c ^\nu +\Gamma _v^\nu $, where
$\Gamma _v^\nu $ is the decay width due to positron detachment
accompanied by the vibrational excitation of the molecule. This leads to
a modification of Eq. (\ref{eq:resres}), which now reads
\begin{equation}\label{eq:ZGv}
Z_{\rm eff}^{\rm (eff)}=\frac{2\pi ^2\rho _{ep}}{k}\rho (E_v)
\left\langle \frac{\Gamma _c^\nu }{\Gamma _c ^\nu +\Gamma _v^\nu }
\right\rangle .
\end{equation}
This equation shows that as soon as the positron energy exceeds another
inelastic vibrational-excitation threshold, the factor in brackets
drops, thereby reducing the resonant annihilation contribution. Such downward
step-like structures at vibrational thresholds are well known in dissociative 
electron attachment experiments (see, e.g., Refs. \cite{Klar:92,Hotop:95}).
When the positron energy is well above the lowest inelastic vibrational
threshold the ``elastic'' width $\Gamma _c$ will become much smaller than the
``inelastic'' width $\Gamma _v$, due to a large number of open
inelastic vibrational-excitation scattering channels, and due to a kinematic
increase of $\Gamma _v$ above the respective thresholds. This will strongly
suppress the resonant annihilation contribution (\ref{eq:ZGv}) with respect to
that of Eq. (\ref{eq:resres}) at larger positron energies. One may speculate
that it is precisely the increase of $\Gamma _v^\nu $ that counteracts the rise
of $\rho (E_v)$, and prevents rapid growth of $Z_{\rm eff}^{\rm (eff)}$
with positron energies. It may also be true that a similar mechanisms is
behind the dramatic drop of the dissociative attachment cross sections for
projectile energies above few lower vibrationally inelastic thresholds
\cite{Christo:84,Klar:92}.

%***************************************************************************
\section{Summary and outlook}\label{sec:concl}

In this work I have considered two possible mechanisms of low-energy positron
annihilation in binary collisions with molecules.

The first mechanisms is direct annihilation. It describes positron annihilation
with atoms and small molecules, as well as molecules which do not form bound
states with the positron. The annihilation rate due to this mechanism has been
related to the positron elastic scattering properties. In particular, it is
enhanced when the positron has a low-lying virtual $s$-type level or a weakly
bound state at $\varepsilon _0=\pm \kappa ^2/2$. For zero-energy positrons
the direct annihilation rate is inversely proportional to $|\varepsilon _0|$.
Small $\kappa $, together with the dipole polarizability of the target,
also determine the rapid energy dependence of $Z_{\rm eff}$ at small positron
energies. Estimates show that for room-temperature positrons $Z_{\rm eff}$
of up to $10^3$ can be produced due the virtual/weakly bound state enhancement.

The second mechanism is resonant annihilation. It is operational when the
positron forms temporary bound states with the molecule. As a necessary
condition, the positron affinity of the molecule must be positive. The
positron capture is a resonant process, whereby the energy of the positron
is transferred into vibrational excitations of the positron-molecule complex.
The contribution of this mechanisms to the annihilation rate is proportional to
the level density of the positron-molecule resonances $\rho $. These resonances
are characterized by the capture width $\Gamma _c$ and annihilation width
$\Gamma _a\sim 1~\mu $eV. For $\Gamma _c>\Gamma _a$ its contribution is
independent of $\Gamma _c$, and is basically determined by the density $\rho $.
The resonant mechanism can give very large annihilation rates (up to $10^8$).
Through its dependence on the vibrational excitation spectrum of the
positron-molecule complex, this mechanism shows high sensitivity to the
chemical composition of the target, and the size of the molecule.
Both are essential features of the experimental data \cite{Iwata:95}.

The difference between the two mechanisms is illustrated most clearly by
comparison of the annihilation rates of alkanes and perfluoroalkanes. For
example, C$_6$H$_{14}$ has $Z_{\rm eff}=120\,000$, whereas for C$_6$F$_{14}$,
$Z_{\rm eff}$ is only 630. The present theory attributes this huge difference
to the fact that perfluorocarbons do not form bound states with the positrons,
and hence, the resonant annihilation is switched off for them. On the
other hand, this mechanism is behind the the high $Z_{\rm eff}$ values of
alkanes.

The experimental group at San Diego has performed a number of measurements
on protonated and deuterated molecules to test the sensitivity of $Z_{\rm eff}$
to the molecular vibrational modes \cite{Iwata:95,Iwata:99}. For example,
their data for benzene show
that a replacement of a single hydrogen atom with deuterium changes the
annihilation rate from $Z_{\rm eff}=15\,000$ for C$_6$H$_6$ to
$Z_{\rm eff}=36\,900$ for C$_6$H$_5$D. On the other hand, the data on
fully protonated vs fully deuterated alkanes shows very little difference
between the two cases. Such behaviour is natural for smaller alkanes,
e.g., methane, where direct annihilation is the dominant mechanism. However,
observed for large alkanes, it cannot be readily interpreted by means of
Eq. (\ref{eq:resres}) or alike. It is possible that the vibrational excitations
are dominated by low-lying C$-$C modes which are weakly affected by
deuteration. On the other hand, deuteration may also influence positron
coupling to the molecular vibrations, which will most likely lead to a
reduction of $\Gamma _c$ in Eq. (\ref{eq:Zres}). If the system is in the regime
where $\Gamma _c\sim \Gamma _a$, this effect may offset the decrease of
the vibrational spacings.

In spite of these difficulties, which could only be resolved by doing
calculations for specific molecules, the present theory offers a consistent
description of positron-molecule annihilation in real terms, through some well
defined parameters which characterize the system. It clearly identifies the
two basic
mechanisms of positron annihilation and discusses their specific features. It
also shows that studies of positron annihilation on molecules may give a unique
insight into the physics of molecular reactions which go through formation of
vibrationally excited intermediate states. Such processes are very likely to
be responsible for large dissociative electron attachment cross sections
observed for molecules such as SF$_6$. They are also of key importance for
the whole class of chemical reactions, namely, for unimolecular reactions
(see, e.g., \cite{Gilbert:90}).

\acknowledgements

This work was strongly stimulated by the vast experimental data of the San
Diego group, and I very much appreciate numerous discussions with its
members, especially C. Surko and K. Iwata. I am thankful to my
colleagues at the University of New South Wales, V. Flambaum, A. Gribakina,
M. Kuchiev, and O. Sushkov for their encouragement and useful discussions.
My thanks also go to S. Buckman for the information on vibrational
excitations and dissociative attachment. Support of my work by the
Australian Research Council is gratefully acknowledged.

%***************************************************************************

%************************************************************************

\begin{table}
\caption{Scattering lengths and fitting parameters for the relation between
$Z_{\rm eff}^{\rm (dir)}$ and $\sigma _{\rm el}$, Eq. (\ref{eq:Zdir1})}
\label{tab:aR}
\begin{tabular}{cccc}
Atom or& $a$ & $R_a$ & $4\pi \rho _e\delta R_a$ \\
molecule & (a.u.) & (a.u.) & (a.u.) \\
\tableline
He & $-$0.52\tablenotemark[1] & 3.9 & 0.21 \\
Ne & $-$0.61\tablenotemark[1] & 5.0 & 0.23 \\
Ar & $-$5.30\tablenotemark[1] & 4.3 & 0.42 \\
Kr & $-$10.4\tablenotemark[1] & 4.2 & 0.41 \\
Xe & $-$45.3\tablenotemark[1] & 4.2 & 0.41 \\
C$_2$H$_4$ & $-$18.5\tablenotemark[2] & 4.4 & 3.0
\end{tabular}
\tablenotetext[1] {Calculated in Ref. \cite{McEachran:79}.}
\tablenotetext[2] {Obtained from the calculations of da Silva {\em et al.}
\cite{daSilva:96}.}
\end{table}

\begin{table}
\caption{Annihilation rates and vibrational frequencies of molecules}
\label{tab:vibfreq}
\begin{tabular}{llrccl}
Molecule & Formula & $Z_{\rm eff}$\tablenotemark[1] & $D$\tablenotemark[2]
(cm$^{-1}$) & Symmetry & Frequencies\tablenotemark[3] (cm$^{-1}$) \\
\tableline
Carbon tetrachloride & CCl$_4$ & 9\,530 & 473 & $A_1$ & 459 \\
Butane & C$_4$H$_{10}$ & 11\,300 & 399 & $A_g$ & 429, 837, 1057, \dots \\
Cyclohexane & C$_6$H$_{12}$ & 20\,000 & 226 &$A_{1g}$& 384, 802, 1158, \dots \\
Pentane & C$_5$H$_{12}$ & 37\,800 & 119 & $A_1$ & 179, 401, 863, \dots \\
Carbon tetrabromide & CBr$_4$ & 39\,800 & 113 & $A_1$ & 269 \\
Hexacloroethane & C$_2$Cl$_6$ & 68\,600 & 65.7 &$A_{1g}$& 164, 431, 976\\
Hexane & C$_6$H$_{14}$ & 120\,000 & 37.6 & $A_g$ & 305, 371, 901, \dots \\
Heptane & C$_7$H$_{16}$ & 242\,000 & 18.6 & $-$ & $-$ 

\end{tabular}

\tablenotetext[1]{Experimental values obtained for room-temperature positrons
in the trap, Ref. \cite{Iwata:95}.}
\tablenotetext[2]{Effective spacing for the resonances in
$Z_{\rm eff}^{\rm (res)}$, Eq. (\ref{eq:ZresT}), corresponding to
experimental data.}
\tablenotetext[3]{Lowest molecular vibrational frequencies of the given
symmetry from Ref. \cite{Sverdlov:74}.}

\end{table}

%************************************************************************

\vspace{-50pt}

\begin{thebibliography}{99}
\bibitem{note} In this paper the word `molecules' is often used interchangeably
with `atoms', or `atoms or molecules'. However, there are specific phenomena,
e.g., low-energy positron capture (Sec. \ref{subsec:res}), which involve
vibrational degrees of freedom, and hence, apply to molecules only.
\bibitem{note1}
The energy of positrons at room temperature, $k_{\rm B}T\approx
25$ meV, is too small to excite any electronic degrees of freedom. The
formation of Ps is also impossible if we consider molecules with ionization
potentials greater than the Ps binding energy of 6.8 eV.
\bibitem{note:channels}
At low positron energy annihilation and emission
are the only decay channels of the resonances. At higher energies
positron emission accompanied by the vibrational (and then electronic)
excitation of the molecule becomes possible, see Sec. \ref{subsec:illres}.
\bibitem{note:pol}
Experimental data for a variety of molecules show that a
non-zero dipole moment of the molecule does not have any direct effect on
$Z_{\rm eff}$, Ref. \cite{Iwata:PhD}. For example, $Z_{\rm eff}=319$, 1090,
and 1600, for H$_2$0, NO$_2$, and NH$_3$, whereas their dipole moments are
1.85, 0.32, and 1.47 Debye, respectively. There also seems to be no
correlation between $Z_{\rm eff}$ values of CH$_3$F, CH$_2$F$_2$, CHF$_3$,
quoted in the text, and their respective dipole moments, 1.85, 1.97, and
1.65 Debye.
\bibitem{note:res}
The contribution of a particular resonance to $Z_{\rm eff}$
is given by the Breit-Wigner formula, cf. terms with $\mu =\nu $ in Eq.
(\ref{eq:zeff1}). For the purpose of thermal averaging, the contribution of
the $\nu $th resonance can be approximated by the $\delta $-function,
$\Delta Z_{\rm eff}= (2\pi ^2/k)\rho _{ep}^{\nu }(\Gamma _c^\nu /\Gamma _\nu )
\delta (E-E_\nu )$, without any loss of accuracy.
\end{thebibliography}
\end{document}